\documentclass{article}
\usepackage{spconf,amsmath,graphicx}
\usepackage{amsmath}
\usepackage{cite}
\usepackage{booktabs}
\usepackage{amssymb}
\usepackage{subcaption}

\newcommand\blfootnote[1]{%
  \begingroup
  \renewcommand\thefootnote{}\footnote{#1}%
  \addtocounter{footnote}{-1}%
  \endgroup
}

\title{Naturalistic Head Motion Generation from Speech}

\twoauthors
{\begin{tabular}{c}
Trisha Mittal{$^{*\dagger}$}
\end{tabular}}
{Department of Computer Science\\University of Maryland, College Park}
{\begin{tabular}{c}
Zakaria Aldeneh$^\dagger$,
Masha Fedzechkina$^\dagger$,\\
Anurag Ranjan,
Barry-John Theobald
\end{tabular}}
{Apple}
\usepackage[citecolor=cyan, colorlinks=true]{hyperref}
\usepackage{enumitem}
\begin{document}
%
\maketitle
\begin{abstract}

Synthesizing natural head motion to accompany speech for an embodied conversational agent is necessary for providing a rich interactive experience. Most prior works assess the quality of generated head motion by comparing them against a single ground-truth using an objective metric. Yet there are many plausible head motion sequences to accompany a speech utterance. In this work, we study the variation in the perceptual quality of head motions sampled from a generative model. We show that, despite providing more diverse head motions, the generative model produces motions with varying degrees of perceptual quality. We finally show that objective metrics commonly used in previous research do not accurately reflect the perceptual quality of generated head motions. These results open an interesting avenue for future work to investigate better objective metrics that correlate with human perception of quality.
\end{abstract}
\section{Introduction}
\label{sec:intro}
Head motion provides a rich source of non-verbal cues in human communication and social interaction. Imbuing AI-based characters and embodied conversational agents with a natural head motion to accompany their speech can lead to a more engaging and immersive interactive experience and an improvement in the intelligibility of the agents' speech \cite{munhall_et_al04_psychsci}. Studies on human interaction suggest that there is a quantifiable relationship between head motion and acoustic attributes~\cite{yehia_et_al02_speech_headmotion}. Consequently, a lot of work focused on the use of machine learning to drive head motion from speech.\blfootnote{$^*$Work done during an internship at Apple.}\blfootnote{$^\dagger$Authors contributed equally.}

Ding et al.~\cite{ding_et_al14_dnn} were the first to successfully use a fully-connected deep neural network to predict head motion from acoustic features. In subsequent work~\cite{ding_et_al15_blstm}, they improved on the previous model by incorporating context using bi-directional long short-term memory (BLSTM) networks. Haag and Shimodaira~\cite{haag2016bidirectional} showed that additional improvement in head motion synthesis can be obtained by appending bottleneck features to the input speech features. While these approaches differ in terms of the proposed model architectures, they are deterministic, i.e., they generate only one head motion sequence for a given a speech signal.

However, the correspondence between speech and head motion is a one-to-many problem; while there is a correlation between the speech signal and head motion, a speaker repeating the same utterance produces different head movements. Thus, deterministic models are unlikely to generate suitably expressive head motions for conversational agents. Recent work has focused on non-deterministic models that can generate more than one head motion trajectory for the same speech. For example, Sadoughi and Busso~\cite{sadoghi_busso18} proposed a conditional generative adversarial network (GAN) that learns a distribution of head motions conditioned on the speech sample and generates a variety of trajectories by sampling from this distribution based on different noise values. Greenwood et al.~\cite{greenwood2017predicting} proposed a conditional variational autoencoder (CVAE) that generates a range of head motion trajectories for the same speech signal by sampling from a Gaussian distribution. While these proposals can produce more varied head motions, to the best of our knowledge, no study has evaluated the quality of the variety of head motions produced by non-deterministic models for the same speech signal. Instead, previous studies either performed subjective evaluations of sample head motion sequences~\cite{sadoghi_busso18} or an informal inspection of the variety in the predicted values rather than a more formal evaluation~\cite{greenwood2017predicting}.

Additionally, objective evaluation of the generated head motions is complicated by the fact that the objective measures used, e.g., mean absolute error (MAE), dynamic time warping (DTW) distance, Pearson correlation coefficient, and the Frechet distance (FD)~\cite{heusel2017gans,li2021learn,ng2022learning}, do not consider what is important in the sense of human perception of naturalness of head motion as they treat all errors equally~\cite{theobald2012relating,hussen2020modality,websdale2021speaker}. Thus, what remains unknown from prior work is whether the variety of head motion sequences produced by non-deterministic models are of consistent perceptual quality.


This work aims to investigate the perceptual quality of head motion sampled from a non-deterministic generative model. We first demonstrate (both qualitatively and quantitatively) that our model generates diverse outputs that contain natural variation in head motion for the same utterance. We then show that the generated head motions vary in their perceptual quality, despite their diversity. We compare against ground-truth sequences and sequences generated by a deterministic baseline model. Finally, we study the relationship between perceptual quality and quantitative metrics (e.g., MAE). Our results provide insights into human perception of head motion and show that measuring the objective quality of head motion sequences does not necessarily align with human perception of quality.

\section{Models}
\label{sec:model}

\subsection{Deterministic baseline model}\label{subsec:deterministic}
We start with a deterministic model, $\mathcal{G}$, that maps a sequence of acoustic features $\mathbf{X}=\{\mathbf{x}_0, \mathbf{x}_1, \dots, \mathbf{x}_T\}$, where $\mathbf{x}_i \in \mathbb{R}^{128}$, to a sequence of head pose parameters $\Theta_h=\{\theta_{h,0}, \theta_{h,1}, \dots, \theta_{h,T}\}$, where
$\theta_h \in SO(3)$,\footnote{$SO(3)$ is the set of rotations in the Euclidean space; a subset of $\mathbb{R}^{3}$.}
such that: $\hat{\Theta}_h = \mathcal{G}(\mathbf{X})$.
The model consists of a multi-layered, bi-directional GRU followed by a fully-connected layer. This architecture was used in multi-modal translation tasks (e.g., gesture generation~\cite{yoon2020speech}). We use an L$1$ loss function for training our deterministic model similar to prior work~\cite{ginosar2019learning,yoon2020speech}. 


\subsection{Non-deterministic model}

To convert the deterministic model described in Section~\ref{subsec:deterministic} to a non-deterministic form, we condition the model on a noise vector $\mathbf{Z}\sim \mathcal{N}(\textbf{0}, \textbf{I}_d)$~\cite{feng2021learning,greshler2021catch}, that is concatenated as an additional channel to the acoustic features such that: $\hat{\Theta}_h = \mathcal{G}(\mathbf{X}, \mathbf{Z})$. The dimensionality of the noise vector matches that of the acoustic features. During training, each input speech utterance is associated with a randomly sampled noise vector. After training, different output sequences for the same speech utterance can be generated by randomly sampling noise vectors and concatenating them to the acoustic features.



\subsection{Model training recipe}
We train the deterministic and non-deterministic models using the Adam optimizer with a learning rate of $0.001$ and train for $50$ epochs. The hyper-parameters we optimize on the validation set include: the number of GRU layers $\{1, 2, 3\}$, the number of features in the hidden state $\{16, 32, 64\}$. Our final models consisted of $2$ GRU layers with $64$ features in the hidden state. We generate $10$ different head motions for each input utterance when testing the non-deterministic model.


\begin{figure}[t]
  \centering
  \includegraphics[width=\columnwidth]{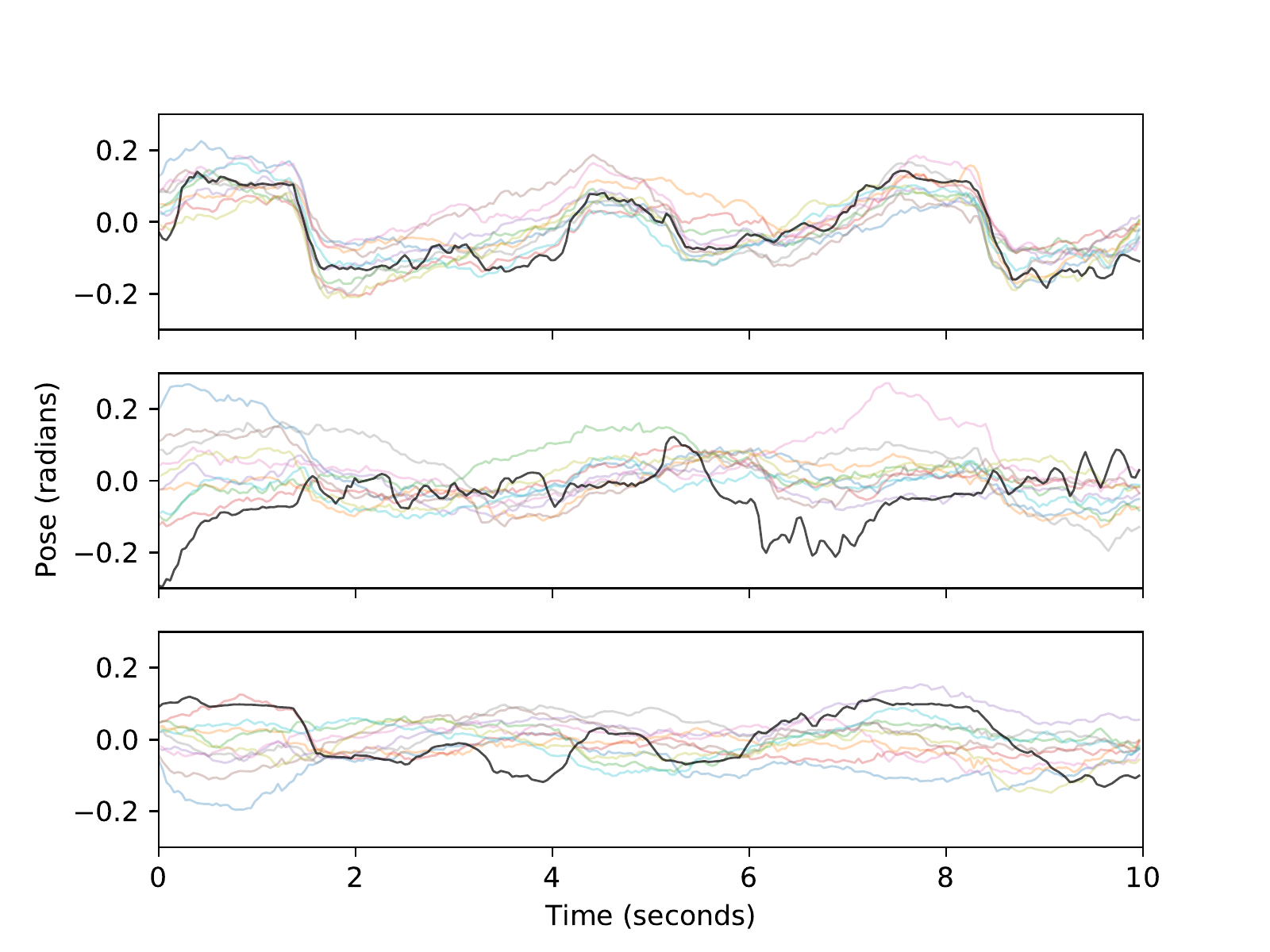}
  \caption{An example of the head motion sequences obtained for a $10$-second test utterance. The top, middle, and bottom figures are for the global rotation head pose parameter across the $x$-, $y$-, and $z$-axis, respectively. The black line represents the ground-truth head motion, and each of the colored lines represents one of the $10$ samples generated using our non-deterministic model.}
  \label{fig:variations_plots}
\end{figure}

\section{Dataset and features}
\label{sec:experiments}
We use the Obama Weekly Address dataset~\cite{suwajanakorn2017synthesizing},  a collection of weekly presidential addresses by former U.S. president Barack Obama. This dataset is commonly used in the talking head generation literature due to its considerable size (${\sim}14$ hours) and consistent camera angle (e.g.,~\cite{lu2021live}). We pre-process the videos in the dataset to have a resolution of $1280\times720$ and a sampling rate of $25$ fps. We split the videos into a training/validation/testing partitions following a $75/12.5/12.5$ strategy. Finally, we further split each video into contiguous 10-second utterances and only retain those from which we can extract head pose parameters for every frame. This process yields $3045$, $503$, and $588$ $10$-second utterances for training, validation, and testing, respectively.

\textbf{3D Morphable Face Model (3DMM)}. We use DECA~\cite{feng2021deca} to estimate the face shape, $\beta$, head pose, $\theta_h$, jaw pose, $\theta_j$ and expression, $\psi$, parameters that drive FLAME 3DMM~\cite{li2017flame}. We use the head pose $\theta_h \in SO(3)$ for our modeling after normalizing the rotation angles by subtracting the mean pose per video.

\textbf{Acoustic Features.} We extract $64$-dimensional Mel-scaled spectrogram features from the $16$ kHz audio using a $40$ ms Hanning window. We $z$-score normalize the frequency bins and use the first- and second-order temporal differences for modeling.


\section{Analysis}

\subsection{Exploring the variability in the generated sequences}
We use the diverseness of predictions measure from~\cite{zhang2020generating,ng2022learning} to quantitatively compare head motions from the proposed non-deterministic model to those from the deterministic model. Specifically, we fit a nearest-neighbor classifier on all training sequences ($n=3045$) and then compute the Shannon index of the nearest-neighbor ID histogram of the predicted motion sequences.
The generated motion in the test sequences  from our proposed non-deterministic model yield a diversity index of $8.48$ while the generated motions from the deterministic model yield a diversity index of $6.62$. The diversity score suggests that non-deterministic models generate more variation in head motion compared deterministic models. Figure~\ref{fig:variations_plots} shows a qualitative example of diverse parameter sequences generated for a random speech utterance from the test set.

\subsection{Exploring the perceptual quality of the head motion sequences}\label{sec:perceptual_quality}
Having established that a non-deterministic model generates a diverse set of head motion sequences for the same utterance, we now ask if the generated motions display variation in terms of their perceptual quality. To this end, we run a user-study to better understand the variability in quality across the $10$ sequences produced by the non-deterministic model for each test utterance.

\subsubsection{The user study}\label{sec:annotation_task}

To assess the perceptual quality of generated head motion sequences, we randomly select $107$ utterances from our test set by randomly selecting $3$ utterances from each test video. For each of these utterances, we generate the head motion sequences with the deterministic model and $10$ head motion sequences with the non-deterministic model. We then render each head motion sequence on the DECA model. 

In the user-study, the annotators were shown two audio-visual sequences that contained the same speech. One of the videos contained ground-truth head motion, while the other video contained generated head motion produced by one of the models (either deterministic or non-deterministic). The annotators were asked to indicate (forced choice) which of the two videos contained more natural head motion with respect to the utterance. Each video pair was rated by $15$ annotators recruited from an in-house crowd-sourcing platform. All annotators were English speakers (determined using a screening process). We average the $15$ binary annotations for each utterance to give a perceptual score that we use in our analysis. The perceptual score takes a value between $0$ and $1$. A perceptual score of $1$, $0$, and $0.5$ indicates that the annotators consistently preferred the ground-truth motion, consistently preferred the generated motion, and preferred both the ground-truth and generated motions equally, respectively.


 \subsubsection{The variability in perceptual quality across generated motions}\label{sec:perceptual_quality}

 \begin{figure}
  \centering
  \hspace*{1px}
  \includegraphics[width=\columnwidth]{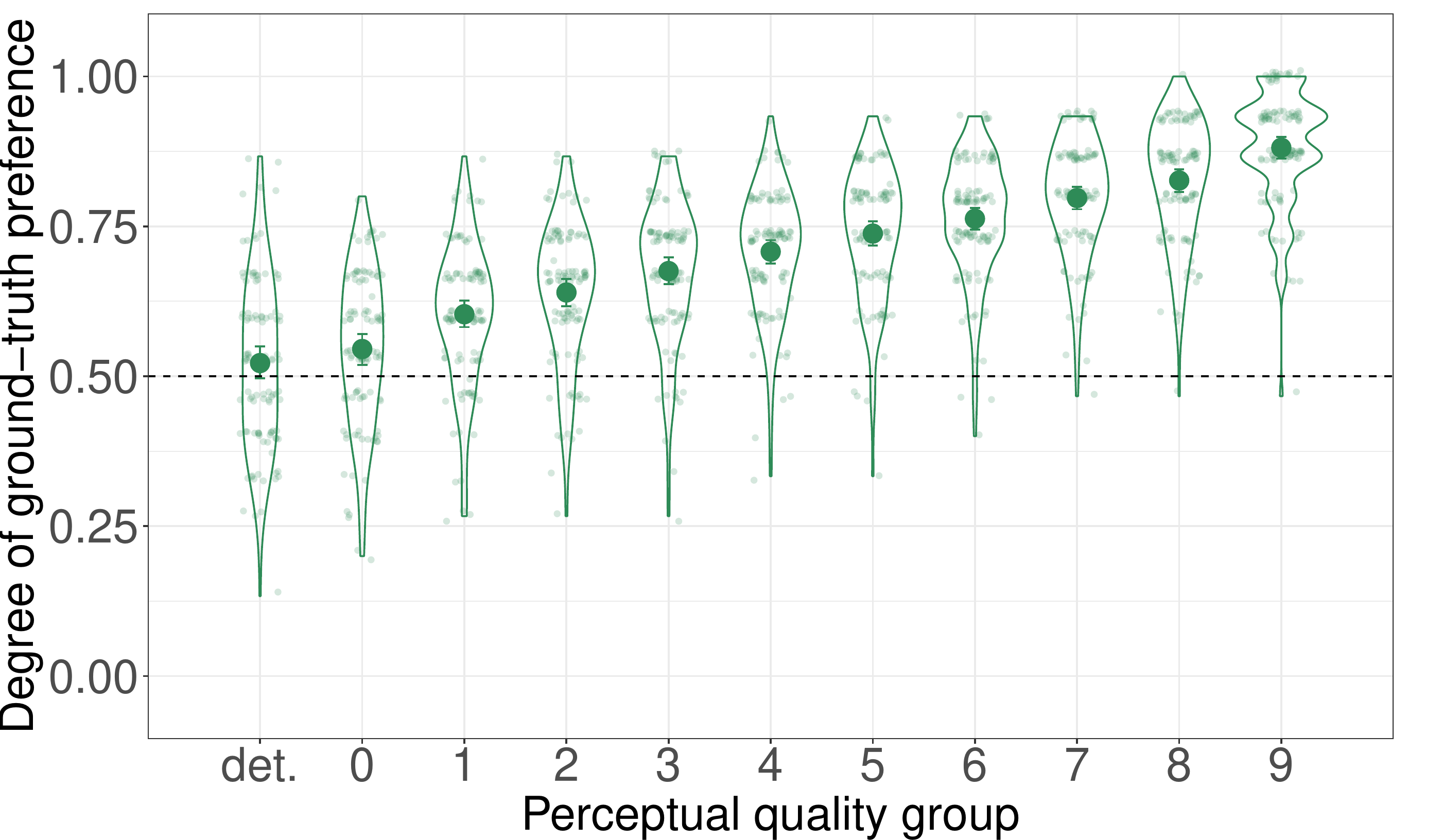}
  \caption{Mean perceptual scores for the head motions produced by the deterministic and non-deterministic models (ranked from best [0] to worst [9]). Large dots represent group means, small dots represent per-utterance means. Error bars represent bootstrapped $95$\% confidence intervals. Perceptual scores that fall on the dashed line (0.5) indicate that the generated head motion is indistinguishable from the ground-truth. Perceptual scores below the dashed line represent generated head motion that is preferred over the ground-truth. Perceptual scores above the dashed line represent generated head motion that is less preferred than the ground-truth.}
  \label{fig:generation_results}
\end{figure}

We rank the $10$ head motions produced by the non-deterministic model for each utterance from most to least natural based on the perceptual scores and then grouped the generated sequences across utterances based on this ranking (so that group $0$ contained the best sequence for each utterance, group $1$ contained the second best sequence, and so on, see Figure~\ref{fig:generation_results}).
As a first step in the analysis, we use mixed-effects logistic regression to compare the perceptual scores given to the sequences generated by the deterministic model and the scores assigned to each perceptual quality group of the non-deterministic model to 0.5 (0.5 indicates no preference between the ground-truth and generated motions). We find no statistically significant difference in the preference for ground-truth sequences and those produced by the deterministic model ($\beta=0.225$, $z=1.53$, $p=0.12$, see Figure~\ref{fig:generation_results}).  However, for all sequences produced by the non-deterministic model, the ground-truth was preferred significantly more (smallest $p=0.03$).

Additional analysis reveals that the best examples produced by the non-deterministic model did not significantly differ from the deterministic model output, suggesting that the sequences produced by both models were perceived with equal quality. However, comparing the perceived quality of the sequences in adjacent groups for the ranked sequences, we find that, with the exception of groups $6$, $8$, and $9$ that did not significantly differ from the previous group, the perceived quality reduced significantly from one group to the next (see Table~\ref{tab:regression_results2}). This result suggests that the sequences produced by the non-deterministic model were of varying perceptual quality---some head motions were indistinguishable from the sequences produced by the deterministic model, whilst others were perceived significantly worse.
\begin{table}[t]
  \centering
\resizebox{\columnwidth}{!}{
  \begin{tabular}{lcccc}
    \toprule
    \textbf{} & \textbf{Estimate} &  \textbf{Std. Error}  &  $\mathbf{z}$ &  $\mathbf{p}$\\
    \midrule
    Group $0$ vs. det.       & $0.04$ & $0.11$ & $0.35$ & $0.720$\\
    Group $1$ vs. $0$       & $0.24$ & $0.08$ & $2.82$ & $0.004$\\
    Group $2$ vs. $1$       & $0.19$ & $0.08$ & $2.82$ & $0.020$\\
    Group $3$ vs. $2$       & $0.21$ & $0.08$ & $2.35$ & $0.010$\\
    Group $4$ vs. $3$     & $0.08$ & $0.09$ & $0.93$ & $0.340$\\
    Group $5$ vs. $4$       & $0.21$ & $0.09$ & $2.31$ & $0.020$\\
    Group $6$ vs. $5$       & $0.09$ & $0.09$ & $1.01$ & $0.300$\\
    Group $7$ vs. $6$       & $0.26$ & $0.09$ & $2.64$ & $0.008$\\
    Group $8$ vs. $7$       & $0.17$ & $0.10$ & $1.67$ & $0.090$\\
    Group $9$ vs. $8$       & $0.43$ & $0.11$ & $3.83$ & $<0.0001$\\
    \bottomrule
  \end{tabular}
  }
  \caption{Mixed-effects logistic regression summary for the analysis of perceptual scores for the head motions produced by the deterministic and non-deterministic models. The model included the maximal random effects structure; group was sliding difference coded (i.e., comparing each subsequent group to the previous group). ``det.'' represents the generations from the deterministic model. The groups represent the ranked ten generations from the non-deterministic model (group 0 contains the best generation for all utterances, group 1 contains the second best generation, etc.). See Section~\ref{sec:perceptual_quality} for model details.}
  \label{tab:regression_results2}
\end{table}

\subsubsection{The relationship between objective metrics and perceptual scores} 
Next, we ask if the perceptual quality of generated head motion correlates with objective metrics commonly used to quantify the quality of head motion. We consider the following objective measures between the ground-truth and predicted head motion: MAE, DTW, and FD. We find that these objective measures are weakly correlated with the perceptual scores (MAE: $r=-0.38$, $p<0.0001$; DTW: $r=-0.39$, $p<0.0001$; FD: $r=-0.38$, $p<0.0001$). These correlations suggest that the perceptual quality of generated head motion cannot be reliably assessed using these objective metrics. Additionally, we find that all three objective measures are highly correlated with each other (MAE and DTW: $r=0.98$, $p<0.0001$; MAE and FD: $r=0.78$, $p<0.0001$), suggesting that including additional objective measures provides little information about head motion realism beyond that already captured by MAE.

\begin{figure}
  \centering
  \includegraphics[width=4cm]{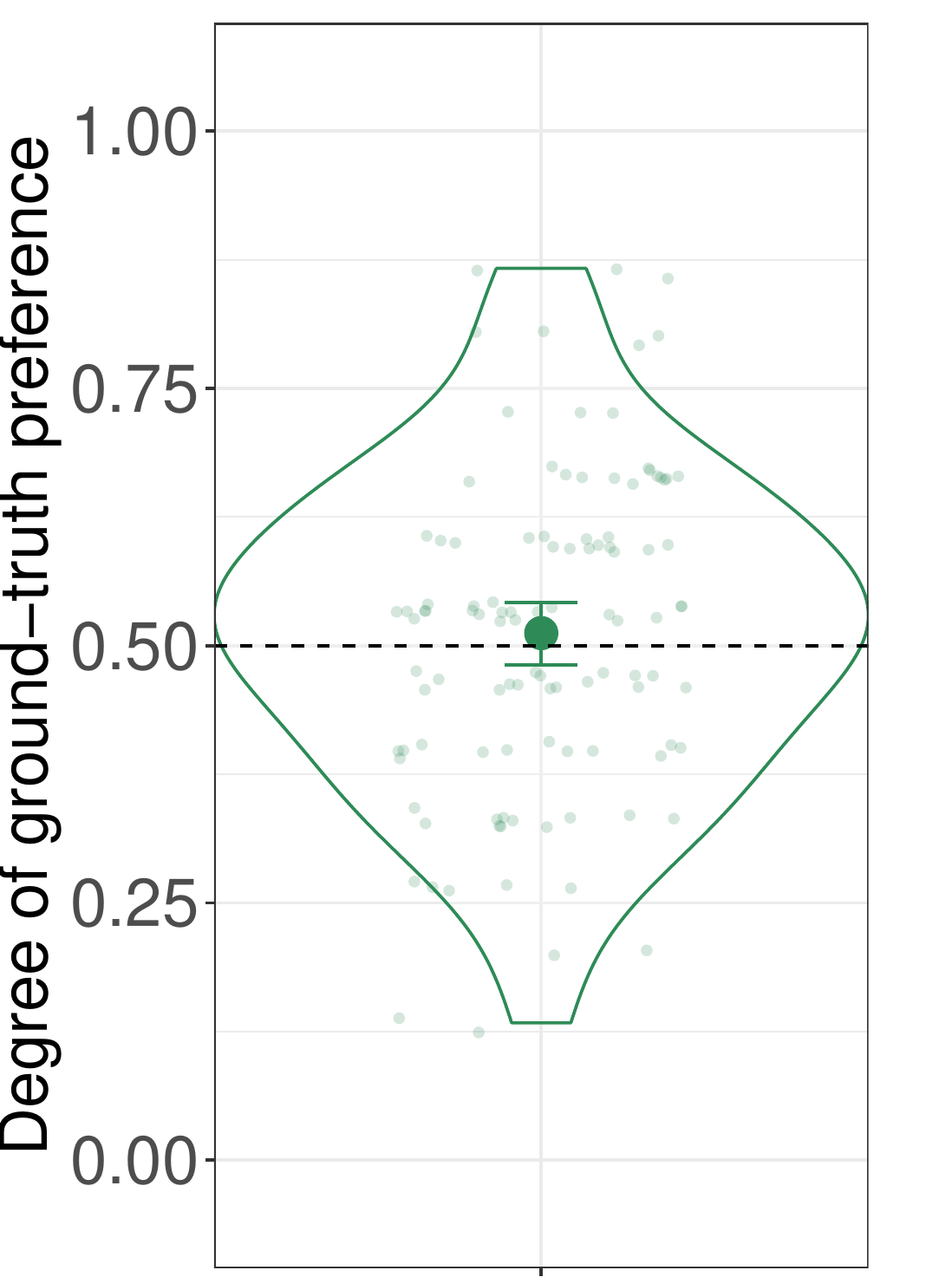}
  \caption{Perceptual scores for mirrored head motion (representation follows Figure~\ref{fig:generation_results})}
  \label{fig:mirror_results}
\end{figure}

  
To better understand the relationship between objective measures and human perception of head motion, we collected perceptual judgments on head motion sequences designed to have a large MAE from the ground-truth. Specifically, for the same $107$ test utterances used in the original user study, we create head motion sequences that are mirrored versions along one of the rotation axes. We then ask a different group of $15$ raters to indicate their preference for the ground-truth or generated head motion using the same forced-choice user-study set-up as in Section~\ref{sec:annotation_task}. Mixed-effects logistic regression reveals that even though there is a relatively large MAE between the mirrored and the ground-truth head motion sequences, there was no statistically significant difference in the perception of the naturalness of these sequences ($\beta=0.005$, $z=0.07$, $p=0.95$, see Figure \ref{fig:mirror_results}). This finding further underscores the discrepancy between commonly used objective metrics, such as MAE, and human perception of naturalness of head motion, suggesting a need for developing objective metrics that better predict human perception.

\section{Conclusion}
We investigated generating realistic head motion from speech. Specifically, we explored the variation in the perceptual quality of head motion sampled from a generative model. We found that while the generative model produces diverse head motion, which is highly desirable for creating naturalistic conversational agents, the head motion had varying degrees of perceptual quality. Thus, an important avenue for future work is to develop a method for ensuring the perceptual quality of the head motion produced by the generative model is consistently high. We also showed that objective metrics, such as MAE, do not accurately reflect the perceptual quality of head motion---a similar finding to that for talking head generation~\cite{theobald2012relating,hussen2020modality,websdale2021speaker}. Our findings underscore the need for future work on developing perceptually-inspired objective metrics that more accurately capture human judgements of quality. Such objective metrics would reduce the need for running costly and time-consuming subjective evaluations via user-studies and could be used as a loss term to train models to better align with viewer preferences.

\textbf{Acknowledgements.}
Thanks to our colleagues, Skyler Seto, Katherine Metcalf, Woojay Jeon, and Russ Webb, for their insightful feedback on this work. Also, thanks to the annotation team and the annotators for their help conducting the perceptual studies.

\bibliographystyle{ieee}
\bibliography{refs}

\begin{thebibliography}{10}

\bibitem{munhall_et_al04_psychsci}
K.G. Munhall, Jefferey~A. Jones, Daniel~E. Callan, Takaaki Kuratate, and Eric
  Vatikiotis-Bateson,
\newblock ``Visual prosody and speech intelligibility: Head movement improves
  auditory speech perception,''
\newblock {\em Psychological Science}, 2004.

\bibitem{yehia_et_al02_speech_headmotion}
Hani~Camille Yehia, Takaaki Kuratate, and Eric Vatikiotis-Bateson,
\newblock ``Linking facial animation, head motion and speech acoustics,''
\newblock {\em Journal of Phonetics}, 2002.

\bibitem{ding_et_al14_dnn}
Chuang Ding, Lei Xie, and Pengcheng Zhu,
\newblock ``Head motion synthesis from speech using deep neural networks,''
\newblock {\em Multimedia Tools and Applications}, 2014.

\bibitem{ding_et_al15_blstm}
Chuang Ding, Pengcheng Zhu, and Lei Xie,
\newblock ``Blstm neural networks for speech driven head motion synthesis,''
\newblock in {\em Interspeech}, 2015.

\bibitem{haag2016bidirectional}
Kathrin Haag and Hiroshi Shimodaira,
\newblock ``Bidirectional lstm networks employing stacked bottleneck features
  for expressive speech-driven head motion synthesis,''
\newblock in {\em International Conference on Intelligent Virtual Agents},
  2016.

\bibitem{sadoghi_busso18}
Najmeh Sadoughi and Carlos Busso,
\newblock ``Novel realizations of speech-driven head movements with generative
  adversarial networks,''
\newblock in {\em IEEE International Conference on Acoustics, Speech and Signal
  Processing (ICASSP)}, 2018.

\bibitem{greenwood2017predicting}
David Greenwood, Stephen Laycock, and Iain Matthews,
\newblock ``Predicting head pose from speech with a conditional variational
  autoencoder,''
\newblock in {\em Interspeech}, 2017.

\bibitem{heusel2017gans}
Martin Heusel, Hubert Ramsauer, Thomas Unterthiner, Bernhard Nessler, and Sepp
  Hochreiter,
\newblock ``Gans trained by a two time-scale update rule converge to a local
  nash equilibrium,''
\newblock {\em Advances in neural information processing systems}, 2017.

\bibitem{li2021learn}
Ruilong Li, Shan Yang, David~A Ross, and Angjoo Kanazawa,
\newblock ``Learn to dance with aist++: Music conditioned 3d dance
  generation,''
\newblock {\em arXiv preprint arXiv:2101.08779}, 2021.

\bibitem{ng2022learning}
Evonne Ng, Hanbyul Joo, Liwen Hu, Hao Li, Trevor Darrell, Angjoo Kanazawa, and
  Shiry Ginosar,
\newblock ``Learning to listen: Modeling non-deterministic dyadic facial
  motion,''
\newblock in {\em Proceedings of the IEEE/CVF Conference on Computer Vision and
  Pattern Recognition}, 2022.

\bibitem{theobald2012relating}
Barry-John Theobald and Iain Matthews,
\newblock ``Relating objective and subjective performance measures for
  aam-based visual speech synthesis,''
\newblock {\em IEEE transactions on audio, speech, and language processing},
  2012.

\bibitem{hussen2020modality}
Ahmed Hussen~Abdelaziz, Barry-John Theobald, Paul Dixon, Reinhard Knothe,
  Nicholas Apostoloff, and Sachin Kajareker,
\newblock ``Modality dropout for improved performance-driven talking faces,''
\newblock in {\em Proceedings of the International Conference on Multimodal
  Interaction}, 2020.

\bibitem{websdale2021speaker}
Danny Websdale, Sarah Taylor, and Ben Milner,
\newblock ``Speaker-independent speech animation using perceptual loss
  functions and synthetic data,''
\newblock {\em IEEE Transactions on Multimedia}, vol. 24, pp. 2539--2552, 2021.

\bibitem{yoon2020speech}
Youngwoo Yoon, Bok Cha, Joo-Haeng Lee, Minsu Jang, Jaeyeon Lee, Jaehong Kim,
  and Geehyuk Lee,
\newblock ``Speech gesture generation from the trimodal context of text, audio,
  and speaker identity,''
\newblock {\em ACM Transactions on Graphics (TOG)}, 2020.

\bibitem{ginosar2019learning}
Shiry Ginosar, Amir Bar, Gefen Kohavi, Caroline Chan, Andrew Owens, and
  Jitendra Malik,
\newblock ``Learning individual styles of conversational gesture,''
\newblock in {\em Proceedings of the IEEE/CVF Conference on Computer Vision and
  Pattern Recognition}, 2019.

\bibitem{feng2021learning}
Yao Feng, Haiwen Feng, Michael~J Black, and Timo Bolkart,
\newblock ``Learning an animatable detailed 3d face model from in-the-wild
  images,''
\newblock {\em ACM Transactions on Graphics (ToG)}, 2021.

\bibitem{greshler2021catch}
Gal Greshler, Tamar Shaham, and Tomer Michaeli,
\newblock ``Catch-a-waveform: Learning to generate audio from a single short
  example,''
\newblock {\em Advances in Neural Information Processing Systems}, 2021.

\bibitem{suwajanakorn2017synthesizing}
Supasorn Suwajanakorn, Steven~M Seitz, and Ira Kemelmacher-Shlizerman,
\newblock ``Synthesizing obama: learning lip sync from audio,''
\newblock {\em ACM Transactions on Graphics (ToG)}, 2017.

\bibitem{lu2021live}
Yuanxun Lu, Jinxiang Chai, and Xun Cao,
\newblock ``Live speech portraits: real-time photorealistic talking-head
  animation,''
\newblock {\em ACM Transactions on Graphics (TOG)}, 2021.

\bibitem{feng2021deca}
Yao Feng, Haiwen Feng, Michael~J Black, and Timo Bolkart,
\newblock ``Learning an animatable detailed 3d face model from in-the-wild
  images,''
\newblock {\em ACM Transactions on Graphics (ToG)}, 2021.

\bibitem{li2017flame}
Tianye Li, Timo Bolkart, Michael~J Black, Hao Li, and Javier Romero,
\newblock ``Learning a model of facial shape and expression from 4d scans.,''
\newblock {\em ACM Transactions on Graphics (TOG)}, 2017.

\bibitem{zhang2020generating}
Yan Zhang, Mohamed Hassan, Heiko Neumann, Michael~J Black, and Siyu Tang,
\newblock ``Generating 3d people in scenes without people,''
\newblock in {\em Proceedings of the IEEE/CVF conference on computer vision and
  pattern recognition}, 2020.

\end{thebibliography}

\end{document}